# The circulation radius and critical current density in type-II superconductors

D.M. Gokhfeld

*Kirensky Institute of Physics, Federal Research Center KSC SB RAS, Krasnoyarsk, Russia*
*Siberian Federal University, Krasnoyarsk, Russia*

A method is proposed for estimating the length scale of currents circulating in superconductors. The estimated circulation radius is used to determine the critical current density from magnetic measurements. The obtained formulas are applicable to samples with negligibly small demagnetizing factors and to polycrystalline superconductors. The proposed method has been verified using experimental magnetization loops measured for polycrystalline $YBa_2Cu_3O_{7-\delta}$ and $Bi_{1.8}Pb_{0.3}Sr_{1.9}Ca_2Cu_3O_x$ superconductors.

The development of cryogenic technology and considerable progress achieved in manufacturing superconducting tapes and single-crystalline samples [1] are disclosing ways to implement superconductors in microelectronics, power engineering, and transport engineering. At the same time, a significant amount of the research devoted to the influence of material structure and/or composition on the properties of superconductors is still performed on polycrystalline samples. This circumstance is related to the relative simplicity of the methods of synthesis and modification of polycrystalline superconductors as compared to single crystals.

In single crystals of high-temperature superconductors (HTSCs), critical current density $J_c$ can reach up to $\sim 10^{12}$ A/m$^2$, which is close to the values of depairing current density [2, 3]. Due to these high values, $J_c$ is usually determined by indirect methods based on magnetic measurements, rather than by direct charge transport measurement techniques. In polycrystalline HTSCs, transport measurements can be used for determining the density of the intergrain critical current, which is several orders of magnitude lower than $J_c$ for single-crystalline samples [2]. Therefore, the intragrain critical current density of polycrystalline HTSCs is always determined using magnetic measurements. The determination and comparison of parameters of various promising superconductors by indirect techniques requires correct data interpretation and taking the particular granular structure into account.

According to the critical state model [4], the magnetization of a type-II superconductor is determined by critical current density $J_c$ and size of the sample. The corresponding expression (Bean's formula) is widely used to find $J_c$ from the results of magnetic measurements as

$$J_c(H) = \Delta M(H)/kR, \qquad (1)$$

where $k$ is the coefficient dependent on the sample geometry ($k = 2/3$ for round or square sample cross section), $R$ is the radius of current circulation (equal to the sample radius), and $\Delta M(H)$ is the magnetization width in external magnetic field $H$. The latter value is defined as $\Delta M(H) = M\downarrow(H) - M\uparrow(H)$, here $M\uparrow(H)$ and $M\downarrow(H)$ are the values the magnetization for the field $H$ increase and decrease respectively. For inhomogeneous superconductors, the circulation radius $R$ in formula (1) can be smaller than the radius of the sample [5, 6] and corresponds to the average value of grain sizes or radius of clusters formed by twin boundaries, cracks, or several jointed grains. The correct estimation of $J_c$ depends on the appropriate choice of $R$. Unjustified use of the sample size as $R$ for polycrystalline superconductors frequently leads to significantly understated $J_c$ values.

A method for determining the circulation radius (current-carrying length) scale was proposed by Angadi et al. [7] for a cylindrical superconducting plate of radius $R$ and thickness $t$ perpendicular to the external field $H$. The demagnetizing factor dependent on the $R/t$ ratio influences the slope of the linear portion of $M\downarrow(H)$ dependence for the field decreasing upon passage through maximum value $H_{max}$. To correctly determine the circulation radius, it is necessary to thoroughly (at small a $H$ step) measure $M\downarrow(H)$ in the region about $M_{max}$. This method is applicable to superconducting single-crystals and films with rather large $R/t$ ratios ($R/t > 2$).

In the present work, a new method is proposed that allows the current circulation radius $R$ to be estimated for samples with a negligibly small demagnetizing factor. The method is based on an analysis of asymmetry of the magnetization loop relative to the axis of field $H$.

Asymmetric magnetization loops of superconductors are described by the extended critical state model [8, 9]. According to this model, it is assumed that a surface region of the sample is characterized by equilibrium magnetization – that is, the Abrikosov vortices are not pinned in a layer of thickness $l_s$. The $l_s$ value increases with the magnetic field and is adopted to be equal to the magnetic field penetration depth $\lambda$. The nonequilibrium magnetization of an inner region of the sample with $R$–$\lambda$ depth is described by the critical state model [4]. Since a nondissipated current is possible provided that Abrikosov vortices are pinned on various defects, only the inner region of the sample is involved in the passage of the supercritical current. For large samples with radius $R \gg \lambda$, the influence of the surface region on the macroscopic critical current density can be ignored. In polycrystalline superconductors and small crystals, this noninvolvement of the near-surface region in the supercritical current passage leads to a decrease in the critical current density averaged over the sample cross section as compared to that for large crystals. This decrease is determined as

$$J_c(H) = J_{cb}(H)\,(1 - l_s(H)/R)^3, \qquad (2)$$

where $J_{cb}$ is the critical current density for the sample of size $R \gg \lambda$. Formula (2) has been derived for a long cylindrical sample ($R \ll t$) with the dependence of the local critical current density on the distance to the cylinder axis [8, 9].

In a superconductor with $R \gg \lambda$, the magnetization loop is almost symmetric relative to the axis of field $H$ and the equality $|M\uparrow(H)| = M\downarrow(H)$ is valid. For this sample, Bean's formula (1) can be written as follows:

$$J_{cb}(H) = 2|M\uparrow(H)|/kR. \qquad (3)$$

In polycrystalline superconductors and small crystals, the equilibrium magnetization of the surface layer leads to a significant asymmetry of the total magnetization loop relative to the axis of field $H$. Due to this asymmetry of the total magnetization loop, the values of $|M\uparrow(H)|$ are greater than those of $|M\downarrow(H)|$ for fields $|H|$ exceeding total penetration field $H_p$. The degree of asymmetry is determined by the $l_s(H)/R$ ratio. Therefore, the current circulation radius influences both $\Delta M$ and the asymmetry of magnetization loop relative to the axis of field $H$. Upon substituting formula (1) for $J_c(H)$ and formula (3) for $J_{cb}(H)$ into Eq. (2), we obtain the following expression:

$$R = l_s(H) / [1 - |\Delta M(H)/2M\uparrow(H)|^{1/3}]. \qquad (4)$$

Since the proportionality of $J_c$ and $\Delta M$ is violated near $H = 0$ [6, 10], the $\Delta M$ values should be determined for some $H > 0$. The optimum variant is the field $H_p$ or the field of maximum diamagnetic response that is somewhat lower than $H_p$. Eventually, the scale of current circulation is suggested to be determined using the following expressions:

$$R = \lambda / [1 - |\Delta M(H_p)/2M\uparrow(H_p)|^{1/3}]. \qquad (5)$$

Here, the values of $\lambda$ near 0 K are known for most superconductors. This value can also be estimated from the reversible region of a magnetization loop by using the London model [11].

Magnetic impurities in the sample lead to sloping of the magnetization loop, thus influencing the estimated $R$ values. The accuracy can be improved by preliminary subtracting additional magnetic contributions from the magnetization loop. The sample porosity does not influence the accuracy of $R$ determination. Formula (5) is applicable to the samples and granules with shapes different from the aforementioned cylinder, provided that the demagnetizing factor remains negligibly small.

It should be noted that a rough estimation of $R$ can be rapidly obtained by taking into account that a significant asymmetry of the loop with respect to the axis of field $H$ is observed for $\lambda/R > 0.1$. Thus, for a loop with pronounced asymmetry of the magnetization loop, the circulation radius is $R < 10\lambda$. This estimation is not valid for defect-free pure superconductors, in which the $l_s$ value can significantly exceed $\lambda$. These materials are free of the pinning centers for Abrikosov vortices. The magnetization curves of defect-free superconductors can exhibit no hysteresis even for large samples.

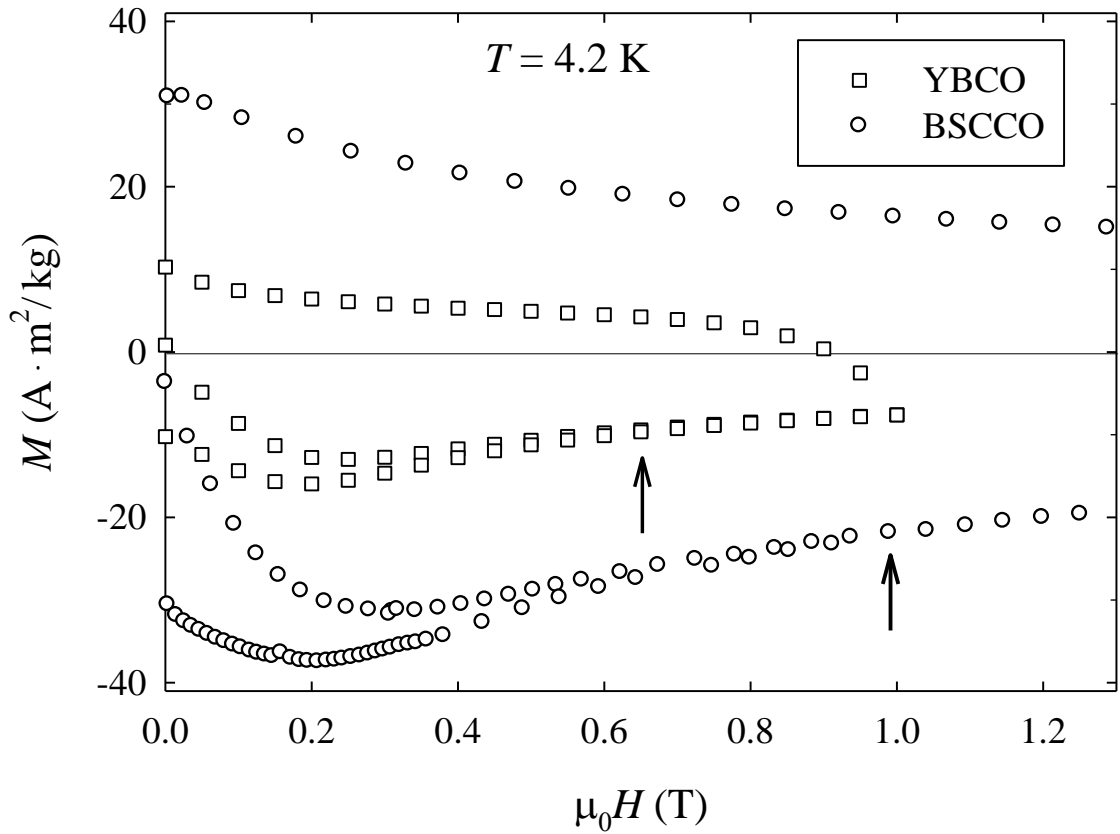

**Fig. 1.** Fragments of magnetization loops $M(H)$ of polycrystalline YBCO and BSCCO samples at 4.2 K. The arrows indicate values of total penetration field $H_p$ for YBCO (~0.65 T) and BSCCO (~0.96 T).

In the present work, formula (5) has been verified for polycrystalline HTSCs. Figure 1 shows the fragments of magnetization curves of polycrystalline $YBa_2Cu_3O_{7-\delta}$ (YBCO) [12] and $Bi_{1.8}Pb_{0.3}Sr_{1.9}Ca_2Cu_3O_x$ (BSCCO) [13] superconductors measured at 4.2 K. The loops are asymmetric relative to the axis of field $H$, the asymmetry being more pronounced for YBCO. The arrows indicate values of total penetration field $H_p$ for YBCO (~0.65 T) and BSCCO (~0.96 T) corresponding to points where the region of initial magnetization merges with the $M\uparrow(H)$ region. From these magnetization loops, it can be determined that $\Delta M(0.65\ T) = 13.9$ A m$^2$/kg and $M\uparrow(0.65\ T) = -9.6$ A m$^2$/kg for YBCO and $\Delta M(0.96\ T) = 38.0$ A m$^2$/kg and $M\uparrow(0.96\ T) = -21.7$ A m$^2$/kg for BSCCO. Assuming $\lambda = 150$ nm for both YBCO and BSCCO [2, 3] and using Eq. (5), we obtain $R \approx 1.5$ μm for YBCO and $R \approx 3.5$ μm for BSCCO. The obtained $R$ values are close to grain radii in the *ab* plane as determined from electron micrographs of the corresponding materials. The critical current density can be estimated by formula (1) as $J_c(0.65\ T) = 8.8\ 10^{10}$ A/m$^2$ for YBCO and $J_c(0.96\ T) = 9.7\ 10^{10}$ A/m$^2$ for BSCCO. The magnetization units (A m$^2$/kg) were converted into current density units (A/m) using theoretical values of the physical density of YBCO and BSSCO. Then, using formulas (2) and (3), the critical current density corresponding to massive single-crystalline samples in low magnetic fields can be estimated as $J_{cb} \approx 9.7 \times 10^{10}$ A/m$^2$ for YBCO and $J_{cb} \approx 10 \times 10^{10}$ A/m$^2$ for BSCCO.

In concluding, the proposed simple and universal method has been developed for estimating the current circulation radius in superconducting samples with negligibly small demagnetizing factor. A comparison of the parameters of promising superconductors determined from the measurements on poly- and single-crystalline superconductors can be performed using Bean's formula in combination with formula (5) for the current circulation radius $R$.

1. M. Miryala and M. R. Koblischka, High-Temperature Superconductors: Occurrence, Synthesis and Applications (Nova Publ., New York, 2018).
2. G. Wang, M. J. Raine, and D. P. Hampshire, Supercond. Sci. Technol. **30**, 104001 (2017).
3. D. Larbalestier, A. Gurevich, D. M. Feldmann, A. Polyanskii, Nature **414**, 368 (2001).
4. C. P. Bean, Rev. Mod. Phys. **36**, 31 (1964).
5. J. Horvat, S. Soltanian, A. V. Pan, and X. L. Wang, J. Appl. Phys. **96**, 4342 (2004).
6. M. Zehetmayer, Phys. Rev. B **80**, 104512 (2009).
7. M. A. Angadi, A. D. Caplin, J. R. Laverty, and Z. X. Shen, Physica C **177**, 479 (1991).
8. D. M. Gokhfeld, Phys. Solid State **56**, 2380 (2014).
9. D. M. Gokhfeld, J. Phys.: Conf. Ser. **695**, 012008 (2016).
10. R. Lal, Physica C **470**, 281 (2010).
11. Z. Hao and J. R. Clem, Phys. Rev. Lett. **67**, 2371 (1991).
12. A. A. Lepeshev, G. S. Patrin, G. Yu. Yurkin, A. D. Vasiliev, I. V. Nemtsev, D. M. Gokhfeld, A. D. Balaev, V. G. Demin, E. P. Bachurina, I. V. Karpov, A. V. Ushakov, L. Yu. Fedorov, L. A. Irtyugo, and M. I. Petrov, J. Supercond. Novel Magn. **31**, 3841 (2018).
13. D. M. Gokhfeld, D. A. Balaev, S. I. Popkov, K. A. Shaykhutdinov, and M. I. Petrov, Physica C **434**, 135 (2006).